\documentclass[12pt]{article}
\usepackage{amsfonts,amssymb,amsmath}
\usepackage{graphicx}
\usepackage{cite}
\usepackage{bm}
\newcommand{\ind}[2]{^{#1}_{\mbox{\scriptsize #2}}}
\newcommand{\ingr}[1]{\centerline{\includegraphics[width=100mm,clip]{#1}}}
\def\nf{n_{\mbox{\scriptsize f}}}

\begin{document}

\begin{center}

{\Large\bf A nonperturbative model for the strong running
coupling within potential approach

}

\vskip10mm

{\large Yu.O.~Belyakova$^{a}$ and A.V.~Nesterenko$^{b}$}

\vskip7.5mm

$^{a}${\small\it Department of Theoretical Physics, Faculty of Physics,\\
Moscow State University, Moscow 119991, Russian Federation}

\vskip5mm

$^{b}${\small\it Bogoliubov Laboratory of Theoretical Physics,\\
Joint Institute for Nuclear Research, Dubna 141980, Russian Federation\\
E--mail:~nesterav@theor.jinr.ru}

\end{center}

\vskip5mm \hrule \vskip4mm

{\bf Abstract}

\vskip0.5mm

A nonperturbative model for the QCD invariant charge, which contains no
low--energy unphysical singularities and possesses an elevated higher loop
corrections stability, is developed in the framework of potential
approach. The static quark--antiquark potential is constructed by making
use of the proposed model for the strong running coupling. The obtained
result coincides with the perturbative potential at small distances and
agrees with relevant lattice simulation data in the nonperturbative
physically--relevant region. The developed model yields a reasonable value
of the QCD scale parameter, which is consistent with its previous
estimations obtained within potential approach.

\vskip2.5mm


\vskip2.5mm


\vskip5mm \hrule \vskip12.5mm
\parindent=7mm

\section{Introduction}

Theoretical description of hadron dynamics at large distances remains a
crucial challenge of elementary particle physics for a long time. The
asymptotic freedom of Quantum Chromodynamics (QCD) allows one to apply
perturbation theory to the study of some ``short--range'' processes, for
example, the high--energy hadronic reactions. However, the study of many
phenomena related to the ``long--range'' dynamics (such as confinement of
quarks, structure of the QCD vacuum, etc.) can be performed only within
nonperturbative methods.

In general, there is a variety of nonperturbative approaches to handle the
strong interaction processes at low energies. For example, one can gain
some hints about the hadron dynamics in infrared domain from lattice
simulations~\cite{QCDLat}, string models~\cite{Nest}, AdS/CFT
methods~\cite{AdSCFT1, AdSCFT2}, sum rules~\cite{OPE1, OPE2}, dispersive
(or~analytic) approach to QCD~\cite{APT, Review, Cvetic}, bag
models~\cite{Bag}, potential models~\cite{Potent, Richardson, BGT, PMRevs,
Lucha, NAWF}. In what follows we shall employ the latter approach, that
involves the construction of the QCD invariant charge which satisfies
certain nonperturbative requirements.

The objective of this paper is to develop a model for the QCD invariant
charge in the framework of potential approach. It is also of a primary
interest to apply the proposed model to the construction of the static
quark--antiquark potential and compare the obtained result with relevant
lattice simulation data.

The layout of the paper is as follows. In Sect.~\ref{Sect:NPIC} the model
for the strong running coupling is formulated and its properties are
discussed. Section~\ref{Sect:Vr} contains the construction of the
potential of quark--antiquark interaction and its comparison with lattice
data. In Conclusions (Sect.~\ref{Sect:Concl}) the obtained results are
summarized and further studies within the approach on hand are outlined.
Appendix~\ref{Sect:BetaPert} contains the explicit expressions for the
perturbative QCD $\beta$--function and strong running coupling up to the
four--loop level. A brief description of the multi--valued Lambert
$W$--function is given in App.~\ref{Sect:Lambert}.

\section{A nonperturbative strong running coupling}
\label{Sect:NPIC}

As it has been mentioned in the Introduction, the asymptotic freedom
enables one to study the strong interaction processes at high energies
within perturbative approach. However, the low--energy hadron dynamics
entirely remains beyond the applicability range of perturbation theory. In
what follows we shall adhere the so--called potential
approach~\cite{Potent, Richardson, BGT, PMRevs, Lucha, NAWF} to the
construction of the nonperturbative model for the QCD invariant charge.

Specifically, in accordance with the basic idea of the potential approach,
we shall construct the strong running coupling~$\alpha(Q^2)$ that
coincides with perturbative QCD invariant charge in the ultraviolet domain
\begin{equation}
\label{AlphaAF}
\alpha(Q^2) \simeq \alpha\ind{}{pert}(Q^2), \qquad Q^2 \to \infty,
\end{equation}
and meets the requirement of the infrared enhancement\footnote{It is worth
noting here that the low--energy behavior of the strong running
coupling~(\ref{AlphaIE}) corresponds to the linearly rising at large
distances static quark--antiquark potential in the framework of the
one--gluon exchange model (see discussion of this issue in
reviews~\cite{PMRevs, Lucha, NAWF}).}
\begin{equation}
\label{AlphaIE}
\alpha(Q^2) \simeq \frac{4\pi}{\beta_0}\,\frac{\Lambda^2}{Q^2},
\qquad Q^2 \to 0_{+}.
\end{equation}
Here $\alpha\ind{}{pert}(Q^2)$ is the perturbative strong running coupling
(see App.~\ref{Sect:BetaPert}), $\beta_0 = 11 - 2\,\nf/3$ denotes the
first coefficient of the $\beta$--function perturbative expansion,
$\nf$~stands for the number of active quarks, $Q^2 \ge 0$ is the spacelike
momentum transferred, and $\Lambda$~denotes the QCD scale parameter. In
terms of the renormalization group $\beta$--function
\begin{equation}
\label{BetaDef}
\frac{d\,\ln a(\mu^2)}{d\,\ln\mu^2} = \beta(a)
\end{equation}
the conditions~(\ref{AlphaAF}) and~(\ref{AlphaIE}) can be equivalently
rewritten as
\begin{equation}
\label{BetaAF}
\beta(a) \simeq \beta\ind{}{pert}(a), \qquad a \to 0_{+},
\end{equation}
and
\begin{equation}
\label{BetaIE}
\beta(a) \simeq -1, \qquad a \to \infty,
\end{equation}
respectively. In these equations $a(Q^2)\equiv\alpha(Q^2)\beta_0/(4\pi)$ stands
for the so--called ``couplant''.

\begin{figure}[t]
\ingr{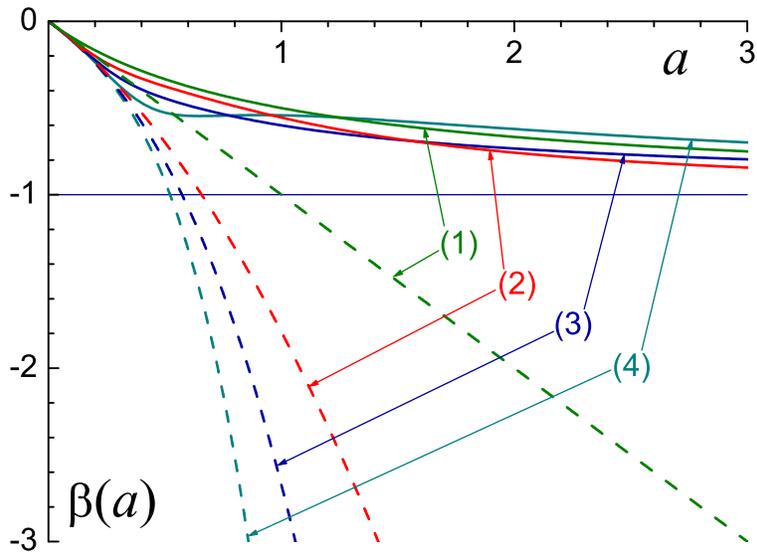}
\caption{The $\beta$--function $\beta^{(\ell)}(a)$~(\ref{BetaLL})
(solid curves) and the perturbative $\beta$--function
$\beta\ind{(\ell)}{pert}(a)$~(\ref{BetaPert}) (dashed curves). The
presented functions correspond to $\nf=3$ active quarks. The numerical
labels indicate the loop level.}
\label{Plot:BetaLL}
\end{figure}

One of the possible expressions for the $\beta$--function that satisfies
conditions~(\ref{BetaAF}) and~(\ref{BetaIE}) reads\footnote{Similar
modifications of the $\beta$--function have been proposed in, e.g.,
Refs.~\cite{Potent, BGT}, whereas an equivalent modification of the
running coupling itself was considered in, e.g., Ref.~\cite{Richardson}.}
\begin{equation}
\label{BetaLL}
\beta^{(\ell)}(a) = \beta\ind{(\ell)}{pert}(a)\,
\frac{1-\exp(-2/a)(1-\ell^2/B_{\ell-1})}{1+\ell^2 a^{\ell}},
\end{equation}
where $\beta\ind{(\ell)}{pert}(a)$ is the $\ell$--loop perturbative QCD
$\beta$--function~(\ref{BetaPert}) and
$B_{n}\!=\!\beta_{n}/\beta_{0}^{n+1}$ is the ratio of the
$\beta$--function expansion coefficients (see App.~\ref{Sect:BetaPert}).
It is straightforward to verify that at any given loop level~$\ell$ the
$\beta$--function~(\ref{BetaLL}) satisfies the aforementioned condition of
linear confinement~(\ref{BetaIE}):
\begin{equation}
\beta^{(\ell)}(a) \simeq -1 + \frac{C_{\ell}}{a} +
{\cal O}\!\left(\frac{1}{a^2}\right)\!,
\qquad a \to \infty,
\end{equation}
where
\begin{equation}
\label{CDef}
C_{\ell} =
\begin{cases}
\:1\,,& \quad \ell=1\\[1.5mm]
\displaystyle
\:2\!\left(1-\frac{\beta_{\ell-1}}{\ell^2 \beta_{0}^{\ell}}\right)-
\beta_{0}\frac{\beta_{\ell-2}}{\beta_{\ell-1}}\,,& \quad \ell>1
\end{cases}
\end{equation}
is the combination of the $\beta$--function expansion coefficients.
At the same time, for small values of the couplant~$a$ the
$\beta$--function~(\ref{BetaLL}) coincides with the corresponding
perturbative result~(\ref{BetaPert}) up to the uncontrollable at
$\ell$--loop level corrections:
\begin{equation}
\beta^{(\ell)}(a) \simeq \beta\ind{(\ell)}{pert}(a) +
{\cal O}\bigl(a^{\ell+1}\bigr),
\qquad a \to 0_{+}.
\end{equation}
Additionally, $\beta$--function~(\ref{BetaLL}) admits the explicit
integration of the renormalization group equation~(\ref{BetaDef}) at the
one--loop level and eventually leads to the strong running
coupling~$\alpha(Q^2)$ that contains no unphysical singularities at low
energies and possesses an elevated stability with respect to the higher
loop corrections. The plots of the
function~$\beta^{(\ell)}(a)$~(\ref{BetaLL}) and the perturbative
$\beta$--function~(\ref{BetaPert}) at $\ell$--loop level ($\ell =
1,...,4$) are presented in Figure~\ref{Plot:BetaLL}.

\begin{figure}[t]
\ingr{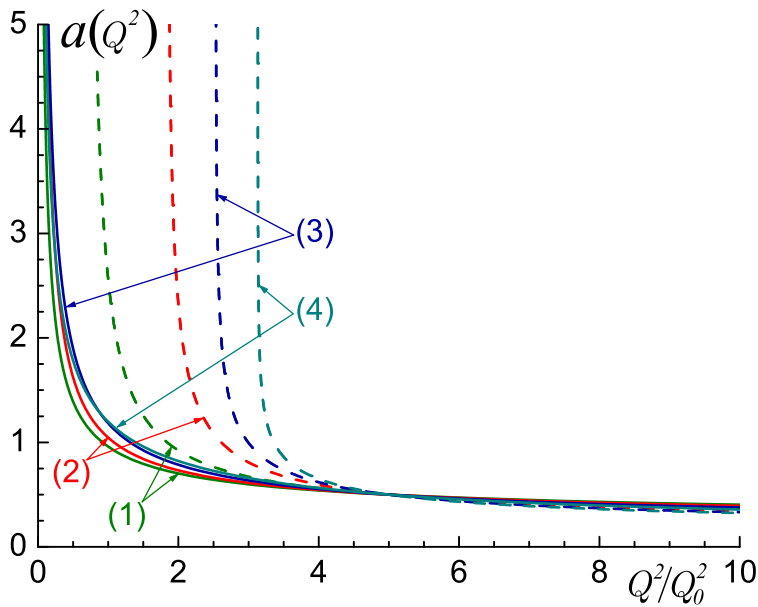}
\caption{The $\ell$--loop couplant~$a^{(\ell)}(Q^2)$ corresponding to the
$\beta$--function~(\ref{BetaLL}) (solid curves) and the perturbative
couplant~$a\ind{(\ell)}{pert}(Q^2)$ (dashed curves). The presented
functions are computed for~$\nf=3$ active quarks and normalized to the
value $a(Q^{2})=1/2$ at~$Q^2=5\,Q_{0}^{2}$. The numerical labels indicate
the loop level.}
\label{Plot:AlphaLL}
\end{figure}

For the beginning, let us consider the one--loop level~($\ell=1$). In this
case the $\beta$--function~(\ref{BetaLL}) takes a simple form, namely,
\begin{equation}
\beta^{(1)}(a) = -\frac{a}{1+a}.
\end{equation}
The corresponding renormalization group equation for the QCD invariant
charge~$\alpha(Q^2)$ reads
\begin{equation}
\label{RGEqn1L}
\frac{d\,\ln\bigl[a^{(1)}(\mu^2)\bigr]}{d\,\ln\mu^2} =
-\frac{a^{(1)}(\mu^2)}{1+a^{(1)}(\mu^2)}.
\end{equation}
After splitting the variables and integrating in finite limits,
Eq.~(\ref{RGEqn1L}) takes the following form:
\begin{equation}
\label{Alpha1LTemp}
\frac{1}{a^{(1)}(Q^2)} - \ln\Bigl[a^{(1)}(Q^2)\Bigr] =
\ln\Biggl(\frac{Q^2}{\Lambda^2}\Biggr),
\end{equation}
where
\begin{equation}
\Lambda^2 = Q_{0}^{2}\,\frac{\beta_{0}}{4\pi}\,\alpha^{(1)}(Q_{0}^{2})
\exp\!\Biggl[-\frac{4\pi}{\beta_{0}}\,\frac{1}{\alpha^{(1)}(Q_{0}^{2})}\Biggr]
\end{equation}
denotes the square of the one--loop scale parameter and $Q_{0}^{2}$ is the
normalization point. In turn, Eq.~(\ref{Alpha1LTemp}) can be solved
explicitly in terms of the multi--valued
Lambert~$W$--function\footnote{The definition of the Lambert~$W$--function
and its properties are described in Ref.~\cite{Lambert} and briefly
overviewed in App.~\ref{Sect:Lambert}.}, namely,
\begin{equation}
\label{Alpha1L}
\alpha^{(1)}(Q^2) = \frac{4\pi}{\beta_0}\,\frac{1}{W_{0}(z)},
\qquad z=\frac{Q^2}{\Lambda^2},
\end{equation}
see also Refs.~\cite{Yulya1, Yulya2}. It is worthwhile to note here that
only the principle branch of the Lambert~$W$--function, $W_{0}(x)$, is
physically meaningful herein. The plots of the couplant corresponding to
the invariant charge~(\ref{Alpha1L}), $a^{(1)}(Q^2)=1/W_{0}(z)$, and the
one--loop perturbative couplant~(\ref{aPert1L}), $a\ind{(1)}{pert}(Q^2)=
1/\ln z$, are presented in Figure~\ref{Plot:AlphaLL} (solid and dashed
curves labeled ``$(1)$'', respectively).

The obtained strong running coupling~(\ref{Alpha1L}) contains no
unphysical singularities in the physical domain~$Q^2>0$. By making use of
the expansions~(\ref{WSeries1}) and~(\ref{WSeries2}) one can show that
$\alpha^{(1)}(Q^2)$~(\ref{Alpha1L}) possesses the required infrared
enhancement,
\begin{equation}
\label{AlphaIE1L}
\alpha^{(1)}(Q^2) \simeq \frac{4\pi}{\beta_0}
\Biggl[\,\frac{1}{z} + 1 + {\cal O}(z) \Biggr],
\qquad Q^2 \to 0_{+},
\end{equation}
and tends to the perturbative result~(\ref{aPert1L}) in the ultraviolet
domain,
\begin{equation}
\alpha^{(1)}(Q^2) \simeq \frac{4\pi}{\beta_0}
\Biggl\{\frac{1}{\ln z} +
{\cal O}\!\Biggl[\frac{\ln(\ln z)}{\ln^{2} z}\Biggr]\!\Biggr\},
\qquad Q^2 \to \infty.
\end{equation}
It is worth mentioning here that the low--energy behavior of the QCD
invariant charge similar to that of Eq.~(\ref{AlphaIE1L}) has also been
discussed in Refs.~\cite{PRD, Papers04, Vienna00, Gargnano02, JPG,
AlekArbu, Raczka, Kiselev}.

Let us proceed now to the higher loop levels~($\ell>1$). Here, the
renormalization group equation~(\ref{BetaDef}) with the
$\beta$--function~(\ref{BetaLL}) can only be integrated numerically.
Nonetheless, the asymptotic behavior of the $\ell$--loop QCD invariant
charge~$\alpha^{(\ell)}(Q^2)$ corresponding to the
$\beta$--function~(\ref{BetaLL}) can be found explicitly. Specifically, at
any loop level~$\alpha^{(\ell)}(Q^2)$ possesses an enhancement in the
infrared asymptotic,
\begin{equation}
\label{AlphaLLIR}
\alpha^{(\ell)}(Q^2) \simeq \frac{4\pi}{\beta_0}
\Biggl[\,\frac{1}{z} + C_{\ell} + {\cal O}(z) \Biggr],
\qquad Q^2 \to 0_{+},
\end{equation}
and coincides with perturbative strong running
coupling~$\alpha\ind{(\ell)}{pert}(Q^2)$ (see App.~\ref{Sect:BetaPert}) in
the ultraviolet asymptotic,
\begin{equation}
\label{AlphaLLUV}
\alpha^{(\ell)}(Q^2) \simeq \alpha\ind{(\ell)}{pert}(Q^2) +
{\cal O}\!\Biggl[\frac{\ln^{\ell}(\ln z)}{\ln^{\ell+1}z}\Biggr],
\qquad Q^2 \to \infty.
\end{equation}
The constant~$C_{\ell}$ in Eq.~(\ref{AlphaLLIR}) is defined in
Eq.~(\ref{CDef}). It is worth noting here that the last term of
Eq.~(\ref{AlphaLLUV}) constitutes a correction, which is not controllable
within perturbative approach at the $\ell$--loop level. The plots of the
couplants~$a^{(\ell)}(Q^2)$ and~$a\ind{(\ell)}{pert}(Q^2)$ at $\ell$--loop
level ($\ell=1,...,4$) are presented in Figure~\ref{Plot:AlphaLL}. As one
can infer from this figure, the QCD invariant
charge~$\alpha^{(\ell)}(Q^2)$ possesses an elevated (with respect to
perturbation theory) higher loop corrections stability in the intermediate
energy range. It is worthwhile to note also that the proposed strong
running coupling~$\alpha(Q^2)$ is free of low--energy unphysical
singularities.

\section{Static quark--antiquark potential}
\label{Sect:Vr}

Let us address now the construction of the static potential of
quark--antiquark interaction~$V(r)$. In the framework of potential
approach~$V(r)$ is related to the strong running coupling $\alpha(Q^2)$,
which satisfies the aforementioned conditions~(\ref{AlphaAF})
and~(\ref{AlphaIE}), by the three--dimensional Fourier transformation
\begin{equation}
\label{VrGen}
V(r) = - \frac{16 \pi}{3} \int\limits_{0}^{\infty}
\frac{\alpha(\bm{Q}^2)}{\bm{Q}^2}\,
\frac{\exp(i \bm{Q r})}{(2 \pi)^3}\, d {\bm Q},
\end{equation}
see, for example, reviews~\cite{PMRevs, Lucha, NAWF} and references
therein. Strictly speaking, this definition of the potential is justified
for small distances ($r \lesssim 0.1\,$fm) only. For instance, the
lowest--lying bound states of heavy quarks can be described by employing
the perturbative\footnote{The leading short--distance nonperturbative
effect due to the gluon condensate has also been accounted for in
Ref.~\cite{Ynd}.} QCD~\cite{Ynd}. However, at large distances ($r \gtrsim
0.5\,$fm), which play the crucial role in hadron spectroscopy, the
perturbative approach becomes inapplicable due to the infrared unphysical
singularities (such as the Landau pole) of the strong running
coupling~$\alpha\ind{}{pert}(Q^2)$. Nonetheless, the
potential~(\ref{VrGen}), being constructed with the invariant
charge~$\alpha(Q^2)$, which contains no unphysical singularities and
satisfies requirements~(\ref{AlphaAF}) and~(\ref{AlphaIE}), has proved to
be successful in the description of both heavy--quark and light--quark
systems (see papers~\cite{Potent, Richardson, BGT} and
reviews~\cite{PMRevs, Lucha, NAWF} for the details).

In this paper, for the construction of the static potential of
quark--antiquark interaction we shall use the invariant
charge~(\ref{Alpha1L}). After integration over angular variables,
Eq.~(\ref{VrGen}) in this case takes the form
\begin{equation}
\label{VrInt}
V(r) = - \frac{32}{3 \beta_0}\,\Lambda\!\int\limits_{0}^{\infty}\!
a(p^2\Lambda^2)\, \frac{\sin (p R)}{p R}\, d p,
\end{equation}
where $p = Q/\Lambda$, $R = r\Lambda$, and
$a(Q^2)=1/W_{0}(Q^2/\Lambda^2)$. The integral~(\ref{VrInt}) diverges at
the lower limit, that is a common feature of the models of such kind (a
detailed discussion of this issue can be found in Sect.~7 and App.~C of
Ref.~\cite{Lucha}). To regularize Eq.~(\ref{VrInt}) it is convenient to
split the couplant~(\ref{Alpha1L}) into singular and regular parts (see
also Eq.~(\ref{AlphaIE1L})):
\begin{equation}
a(Q^2) = a_{1}(z) + a_{2}(z),
\end{equation}
where $z=Q^2/\Lambda^2$ and
\begin{equation}
\label{A1A2Def}
a_{1}(z) = \frac{1}{z}, \qquad
a_{2}(z) = \frac{1}{W_{0}(z)} - \frac{1}{z}.
\end{equation}
Then, the potential~(\ref{VrInt}) takes the following form:
\begin{equation}
\label{VrTemp}
V(r) = \frac{8\pi}{3\beta_0}\,\Lambda
\biggl[U_{1}(R) + U_{2}(R)\biggr],
\end{equation}
where
\begin{equation}
\label{UDef}
U_{i}(R) = - \frac{1}{R}\,\frac{4}{\pi} \int\limits_{0}^{\infty}\!
a_{i}\Biggl(\frac{x^2}{R^2}\Biggr)\frac{\sin x}{x}\, d x,
\qquad i=1,2
\end{equation}
are the dimensionless parts of the potential and~$x=Qr$. The
function~$U_{1}(R)$ diverges and requires regularization,
whereas~$U_{2}(R)$ is regular and can be computed numerically.

To regularize function~$U_{1}(R)$ we shall employ the method similar to
that of used in Ref.~\cite{PRD}. Specifically, in terms of auxiliary
function
\begin{equation}
\label{Aux1}
I(t) = \int\limits_{0}^{\infty}x^{t} \sin x \, d x
\end{equation}
the singular part of the potential~(\ref{VrTemp}) reads
\begin{equation}
\label{U1Sing}
U_{1}(R) = - \frac{4}{\pi}\, I(-3)\,R\,.
\end{equation}
The integral on the right hand side of Eq.~(\ref{Aux1}) exists for $0 \le
|\mbox{Re}\,(t+1)| < 1$ (see, for example, Ref.~\cite{RG}). Nonetheless,
Eq.~(\ref{Aux1}) can be analytically continued to the
entire\footnote{Except for the singular points of the right hand side of
Eq.~(\ref{Aux2}), such as $t = -2N$, with $N=1,2,3,...$~being a natural
number.} complex $t$--plane. This continuation is given by
\begin{equation}
\label{Aux2}
I(t) = \sqrt{\pi}\, 2^t\,
\frac{\Gamma(1+t/2)}{\Gamma\bigl((1 - t)/2\bigr)}
\end{equation}
and plays the role of regularization of Eq.~(\ref{U1Sing}), see also
Refs.~\cite{PRD, Review}. In this case $I(-3) = -\pi/4$, that results
in~$U_{1}(R)=R$.

Thus, the static quark--antiquark potential~(\ref{VrInt}) takes the
following form:
\begin{equation}
\label{Vr}
V(r) = V_{0} + \frac{8\pi}{3\beta_0}\,\Lambda
\Biggl[ R - \frac{1}{R}\,\frac{4}{\pi}
\int\limits_{0}^{\infty}\!
a_{2}\Biggl(\frac{x^2}{R^2}\Biggr)\frac{\sin x}{x}\, d x
\Biggr], \qquad R=\Lambda r,
\end{equation}
where $V_{0}$ is an additive self--energy constant and~$a_{2}(z)$ is given
by~Eq.~(\ref{A1A2Def}). At small distances this potential possesses the
standard behavior determined by the asymptotic freedom
\begin{equation}
\label{VrPert}
V(r) \simeq \frac{8 \pi}{3 \beta_0}\, \Lambda\,
\frac{1}{R\, \ln R}, \qquad r \to 0,
\end{equation}
whereas at large distances potential~(\ref{Vr}) proves to be linearly
rising
\begin{equation}
V(r) \simeq \frac{8\pi}{3\beta_0}\,\Lambda\, R,
\qquad r\to\infty,
\end{equation}
implying the confinement of quarks. It is straightforward to verify that
the potential~(\ref{Vr}) satisfies also the so--called concavity condition
\begin{equation}
\frac{d\,V(r)}{d\,r} > 0, \qquad \frac{d^2\, V(r)}{d\, r^2} \le 0,
\end{equation}
which is a general property of the gauge theories (see Ref.~\cite{SB} for
the details).

\begin{figure}[t]
\ingr{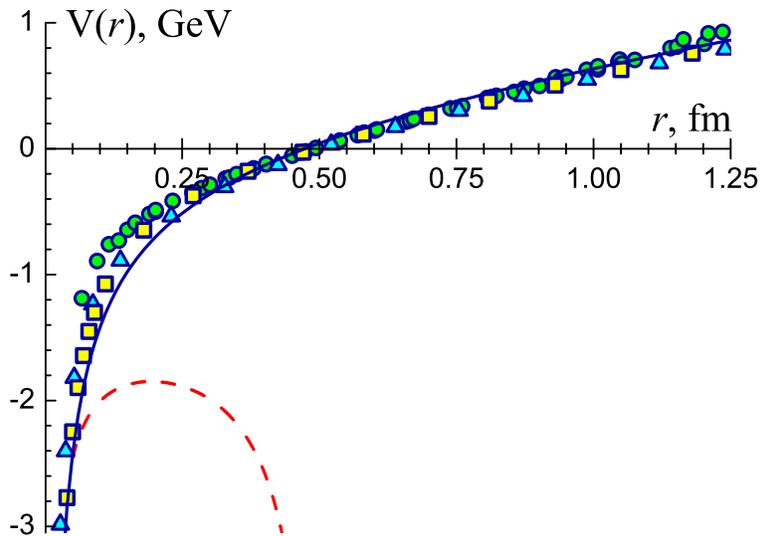}
\caption{The quark--antiquark potential $V(r)$ (Eq.~(\ref{Vr}), solid
curve) and the lattice simulation data (Ref.~\cite{Bali2000},
symbols~``{\protect\raisebox{1.5pt}{{\scriptsize$\bigcirc$}}}'').
The Cornell potential~\cite{Cornell} is denoted by symbols~``$\Box$'',
whereas Richardson's potential~\cite{Richardson} is shown by
symbols~``$\triangle$''. The dashed curve denotes the relevant
perturbative result~(\ref{VrPert}). The values of the parameters:
$\Lambda=375\,$MeV, $V_{0}=315\,$MeV, $\nf=3$.
\label{Plot:Vr}}
\end{figure}

Figure~\ref{Plot:Vr} presents the quark--antiquark potential~(\ref{Vr}),
relevant lattice simulation data~\cite{Bali2000}, Cornell
potential~\cite{Cornell}, and Richardson's potential~\cite{Richardson}.
Equation~(\ref{Vr}) has been fitted to the lattice data~\cite{Bali2000} by
making use of the least square method, $\Lambda$ and~$V_{0}$ being the
varied parameters. The estimation of the scale parameter in the course of
this comparison gives $\Lambda = (375 \pm 40)\,$MeV (this value
corresponds to the one--loop level with $\nf=3$ active quarks). The
obtained value of~$\Lambda$ agrees with previous estimations~\cite{Potent,
Richardson, BGT} of the QCD scale parameter within potential approach. As
one can infer from Figure~\ref{Plot:Vr}, in the region $r \lesssim
0.05\,\mbox{fm}$ the derived potential~(\ref{Vr}) coincides with the
perturbative\footnote{The applicability range of the perturbative
quark--antiquark potential has been discussed in Refs.~\cite{PeterHMP,
BGT}.} result~(\ref{VrPert}). At the same time, $V(r)$~(\ref{Vr}) is in a
good agreement with both Cornell~\cite{Cornell} and
Richardson's~\cite{Richardson} potentials. Besides, in the nonperturbative
physically--relevant range $0.3\,\mbox{fm} \lesssim r \lesssim
1.2\,\mbox{fm}$, in which the average quark separations $\sqrt{\langle
r^2\rangle}$ for quarkonia sits~\cite{Bali9799}, the obtained
potential~(\ref{Vr}) reproduces the lattice data\cite{Bali2000} fairly
well.

\section{Conclusions}
\label{Sect:Concl}

A nonperturbative model for the QCD invariant charge is developed in the
framework of potential approach. The proposed strong running coupling is
free of low--energy unphysical singularities and embodies the asymptotic
freedom with infrared enhancement in a single expression. The model on
hand possesses an elevated (with respect to perturbation theory) higher
loop corrections stability in the intermediate energy range. The static
quark--antiquark potential is constructed by making use of the proposed
model for the strong running coupling. The obtained result coincides with
the perturbative potential at small distances and agrees with relevant
lattice simulation data in the nonperturbative physically--relevant
region. The developed model yields a reasonable value of the QCD scale
parameter, which agrees with its previous estimations obtained in the
framework of potential approach.

In further studies it would undoubtedly be interesting to compute the
meson spectrum by making use of the derived quark--antiquark potential as
well as to apply the developed model for the QCD invariant charge to the
analysis of other strong interaction processes.

\section*{Acknowledgments}

Authors are grateful to Prof.~G.S.~Bali for providing the relevant lattice
simulation data and fruitful discussions and to Prof.~V.I.~Yukalov for
useful comments.

\vspace*{2.5mm}

\appendix

\renewcommand{\theequation}{\thesection.\arabic{equation}}
\setcounter{equation}{0}

\section{The QCD $\beta$--function and running coupling \\
within perturbation theory}
\label{Sect:BetaPert}

The QCD invariant charge~$\alpha(Q^2)$ is the solution of the
renormalization group equation
\begin{equation}
\label{RGDef}
\frac{d\,\ln\bigl[a(\mu^2)\bigr]}{d\,\ln \mu^2} = \beta(a),
\end{equation}
where $a(Q^2)\equiv\alpha(Q^2) \beta_{0}/(4\pi)$ is the so--called
``couplant''. For small values of the running coupling the right hand side
of Eq.~(\ref{RGDef}) is usually approximated by the power series, namely,
\begin{equation}
\label{BetaPert}
\beta(a) \simeq \beta\ind{(\ell)}{pert}(a) =
- \sum_{n=0}^{\ell-1} B_n
\Bigl[a\ind{(\ell)}{}(\mu^2)\Bigr]^{n+1},
\qquad a \to 0_{+},
\end{equation}
where $\ell = 1,2,3,...$ denotes the loop level and
$B_n=\beta_{n}/\beta_{0}^{n+1}$ is the ratio of the
QCD $\beta$--function perturbative expansion
coefficients:
\begin{eqnarray}
\label{BetaPert1L}
\beta_{0} &=& 11 - \frac{2}{3}\,\nf, \\[2.5mm]
\label{BetaPert2L}
\beta_{1} &=& 102 - \frac{38}{3}\,\nf, \\[2.5mm]
\label{BetaPert3L}
\beta_{2} &=& \frac{2857}{2} - \frac{5033}{18}\,\nf +
              \frac{325}{54}\,\nf^2, \\[2.5mm]
\label{BetaPert4L}
\beta_{3} &=& \frac{149753}{6} + 3564\,\zeta(3) -
\left[\frac{1078361}{162}+\frac{6508}{27}\,\zeta(3)\right]\!\nf
\nonumber \\[2.5mm] &&
+ \left[\frac{50065}{162}+\frac{6472}{81}\,\zeta(3)\right]\!\nf^{2} +
\frac{1093}{729}\,\nf^{3}.
\end{eqnarray}
In these equations $\nf$ stands for the number of active quarks and
$\zeta(x)$ denotes the Riemann $\zeta$--function, $\zeta(3) \simeq 1.202$
(see, for example, Ref.~\cite{RG}). The one-- and two--loop coefficients
($\beta_{0}$~and~$\beta_{1}$) are scheme--independent, whereas the
expressions given for~$\beta_{2}$ and~$\beta_{3}$ are calculated in the
$\overline{\mbox{MS}}$~subtraction scheme (see papers~\cite{BetaPert1L,
BetaPert2L, BetaPert3L, BetaPert4L} and references therein for the
details). The plots of the $\beta$--function~(\ref{BetaPert}) are
presented in Figure~\ref{Plot:BetaLL} (dashed curves).

The renormalization group equation corresponding to the perturbative
$\beta$--function (\ref{BetaPert}),
\begin{equation}
\label{RGEqnPert}
\frac{d\,\ln\bigl[a\ind{(\ell)}{pert}(\mu^2)\bigr]}{d\,\ln \mu^2} =
\beta\ind{(\ell)}{pert}(a),
\end{equation}
can be solved explicitly at one-- and two--loop levels ($\ell=1,2$),
namely,
\begin{eqnarray}
\label{aPert1L}
a\ind{(1)}{pert}(Q^2) &=&
\frac{1}{\ln z}, \qquad z=\frac{Q^2}{\Lambda^2}, \\[7mm]
\label{aPert2L}
a\ind{(2)}{pert}(Q^2) &=&
- \frac{B_{1}^{-1}}{1 + W_{-1}\Bigl\{-\exp\Bigl[-\bigl(1 + B_{1}^{-1}\ln z\bigr)
\Bigr]\!\Bigr\}} \nonumber \\[2.5mm]
&\simeq& \frac{1}{\ln z}-B_{1}\frac{\ln(\ln z)}{\ln^{2} z},
\qquad Q^2\to\infty,\\[-2.5mm] \nonumber
\end{eqnarray}
see also Ref.~\cite{Alpha2LW}. In Eq.~(\ref{aPert2L}) $W_{k}(x)$ stands
for the so--called Lambert $W$--function (see App.~\ref{Sect:Lambert} for
the details). Starting from the three--loop level, the exact solution to
the perturbative renormalization group equation~(\ref{RGEqnPert}) can not
be expressed in terms of known functions. Nonetheless, for $\ell \ge 3$
Eq.~(\ref{RGEqnPert}) can be solved iteratively, that eventually leads to
\begin{eqnarray}
\label{aPert3L}
a\ind{(3)}{pert}(Q^2) &\simeq& \frac{1}{\ln z}-B_{1}\frac{\ln(\ln z)}{\ln^{2} z}
\nonumber \\[2.5mm] &&
+\frac{1}{\ln^{3}\!z}\biggl\{B_{1}^{2} \Bigl[\ln^2(\ln z)
- \ln(\ln z)-1\Bigr]+B_{2}\biggr\}, \\[7mm]
\label{aPert4L}
a\ind{(4)}{pert}(Q^2) &\simeq& \frac{1}{\ln z} - B_{1} \frac{\ln(\ln z)}{\ln^{2} z}
\nonumber \\[2.5mm] &&
+\frac{1}{\ln^{3}\!z}\biggl\{B_{1}^{2} \Bigl[\ln^2(\ln z)
-\ln(\ln z)-1\Bigr]+B_{2}\biggr\}
\nonumber \\[2.5mm] &&
+\frac{1}{\ln^{4}\!z}\Biggl\{B_{1}^{3} \biggl[-\ln^{3}(\ln z)
+\frac{5}{2}\ln^{2}(\ln z)
\nonumber \\[2.5mm] &&
+ 2 \ln(\ln z) -\frac{1}{2}\biggr]
- 3B_{1}B_{2}\ln(\ln z) + \frac{1}{2}B_{3}\Biggr\}.\hspace{5mm}
\\[-2.5mm] \nonumber
\end{eqnarray}
It is worth noting here that the presented in Figure~\ref{Plot:AlphaLL}
plots of the $\ell$--loop perturbative couplant~$a\ind{(\ell)}{pert}(Q^2)$
(dashed curves) correspond to exact solutions of the renormalization group
equation~(\ref{RGEqnPert}).

\setcounter{equation}{0}

\section{The Lambert $W$--function}
\label{Sect:Lambert}

\begin{figure}[h]
\centerline{\includegraphics[width=85mm,clip]{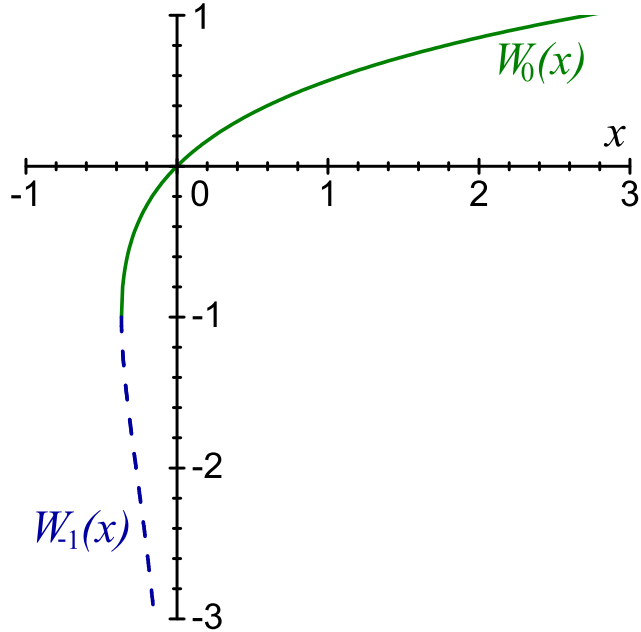}}
\caption{Two real branches of the Lambert
$W$--function~(\ref{LambertDef}): $W_{0}(x)$
(solid curve) and $W_{-1}(x)$ (dashed curve).}
\label{Plot:Lambert}
\end{figure}

The Lambert $W$--function is defined as the multi--valued function
$W_{k}(x)$, which satisfies the equation
\begin{equation}
\label{LambertDef}
W_{k}(x) \exp\Bigl[W_{k}(x)\Bigr] = x,
\end{equation}
where $k$ denotes the branch index. Only two branches of this function,
namely, $W_0(x)$ (the principal branch) and~$W_{-1}(x)$, take real values
(see Figure~\ref{Plot:Lambert}), whereas the other branches take imaginary
values. For the branches $W_0(x)$ and $W_{-1}(x)$ of the Lambert
$W$--function the following expansions hold:
\begin{eqnarray}
\label{WSeries1}
W_{0}(x) &=& \ln x - \ln(\ln x) +
{\cal O}\!\Biggl[\frac{\ln(\ln x)}{\ln x}\Biggr],
\qquad x \to \infty, \qquad~\\[2.5mm]
\label{WSeries2}
W_{0}(x) &=& x - x^2 + {\cal O}(x^3), \qquad x \to 0, \\[2.5mm]
W_{-1}(-x) &=& \ln x + {\cal O}\bigl(\ln |\ln x|\bigr),
\qquad x \to 0_{+}, \\[2.5mm]
W_{-1}\left(-\frac{1}{e} + x \right) &=& -1 - \sqrt{2\,e\,x} + {\cal O}(x),
\qquad x \to 0_{+},
\end{eqnarray}
where $e \simeq 2.71828$ denotes the base of natural logarithm. The
detailed description of the Lambert $W$--function and its properties can
be found in Ref.~\cite{Lambert}.

\end{document}